\RequirePackage{lineno}
\documentclass[aps,twocolumn,showpacs,byrevtex,prl,reprint]{revtex4-1}

\usepackage{graphicx}
\usepackage{dcolumn}
\usepackage{bm}
\usepackage{rotating}
\usepackage{epstopdf}
\usepackage{color}
\usepackage[dvipsnames]{xcolor}
\usepackage{verbatim} 
\usepackage{multirow}
\usepackage[abs]{overpic}
\usepackage{amsmath}
\usepackage{dcolumn}
\usepackage{amssymb}
\usepackage{subfigure}
\usepackage{xspace}
\usepackage{float}
\usepackage{overpic}
\usepackage{lineno}
\usepackage{subfigure}
\usepackage{graphicx}
\usepackage[colorlinks,
            linkcolor=blue,
            anchorcolor=blue,
           citecolor=blue]{hyperref}{}

\usepackage{xcolor}
\usepackage{setspace}
\usepackage{array}

\RequirePackage{lineno}

\usepackage{amsmath}
\newcommand{\PreserveBackslash}[1]{\let\temp=\\#1\let\\=\temp}
\newcolumntype{C}[1]{>{\PreserveBackslash\centering}p{#1}}
\newcolumntype{R}[1]{>{\PreserveBackslash\raggedleft}p{#1}}
\newcolumntype{L}[1]{>{\PreserveBackslash\raggedright}p{#1}}
\newcommand{\pp}{\pi^+\pi^-}
\newcommand{\kk}{K^+K^-}

\newcommand{\EE}{e^+e^-}

\newcommand{\ds}{D_{s}}

\newcommand{\dsstar}{D_{s}^{*}}

\newcommand{\ddstar}{D_{s}^{\pm}D_{s}^{*\mp}}
\newcommand{\dstardstar}{D_{s}^{*\pm}D_{s}^{*\mp}}

\newcommand{\kshort}{K^{0}_{S}}
\newcommand{\mev}{\mathrm{MeV}}
\newcommand{\gev}{\mathrm{GeV}}
\newcommand{\gevcc}{\mathrm{GeV}/c^{2}}
\newcommand{\mevcc}{\mathrm{MeV}/c^{2}}

\def\gev   {\ensuremath{\mbox{\,GeV}\xspace}}

\begin{document}
\graphicspath{{figure/}}
\DeclareGraphicsExtensions{.eps,.png,.ps}
\title{Search for $C$-even states decaying to $D_{s}^{\pm}D_{s}^{*\mp}$ with masses between $4.08$ and $4.32$ $\rm GeV/{\it c}^{2}$ }
\author{
  \begin{small}
    \begin{center}
    M.~Ablikim$^{1}$, M.~N.~Achasov$^{4,c}$, P.~Adlarson$^{75}$, O.~Afedulidis$^{3}$, X.~C.~Ai$^{80}$, R.~Aliberti$^{35}$, A.~Amoroso$^{74A,74C}$, Q.~An$^{71,58,a}$, Y.~Bai$^{57}$, O.~Bakina$^{36}$
% ...
\\
\vspace{0.2cm}
(BESIII Collaboration)\\
\vspace{0.2cm} {\it
$^{1}$ Institute of High Energy Physics, Beijing 100049, People's Republic of China\\
% ...
$^{3}$ Bochum  Ruhr-University, D-44780 Bochum, Germany\\
$^{4}$ Budker Institute of Nuclear Physics SB RAS (BINP), Novosibirsk 630090, Russia\\
% ...
$^{35}$ Johannes Gutenberg University of Mainz, Johann-Joachim-Becher-Weg 45, D-55099 Mainz, Germany\\
$^{36}$ Joint Institute for Nuclear Research, 141980 Dubna, Moscow region, Russia\\
% ...
$^{57}$ Southeast University, Nanjing 211100, People's Republic of China\\
$^{58}$ State Key Laboratory of Particle Detection and Electronics, Beijing 100049, Hefei 230026, People's Republic of China\\
% ...
$^{71}$ University of Science and Technology of China, Hefei 230026, People's Republic of China\\
% ...
$^{74}$ University of Turin and INFN, (A)University of Turin, I-10125, Turin, Italy; (B)University of Eastern Piedmont, I-15121, Alessandria, Italy; (C)INFN, I-10125, Turin, Italy\\
$^{75}$ Uppsala University, Box 516, SE-75120 Uppsala, Sweden\\
% ...
$^{80}$ Zhengzhou University, Zhengzhou 450001, People's Republic of China\\
\vspace{0.2cm}
$^{a}$ Deceased\\
% ...
$^{c}$ Also at the Novosibirsk State University, Novosibirsk, 630090, Russia\\
% ...
}\end{center}
    \vspace{0.4cm}
\end{small}
}
\affiliation{}

\vspace{0.2cm}
\date{\today}

\begin{abstract}
Six $C$-even states, denoted as $X$, with quantum numbers $J^{PC}=0^{-+}$, $1^{\pm+}$, or $2^{\pm+}$, are searched for via 
the $e^+e^-\to\gamma\ddstar$ process using $(1667.39\pm8.84)~\mathrm{pb}^{-1}$ of $e^+e^-$ collision data collected with the 
BESIII detector operating at the BEPCII storage ring at center-of-mass energy of $\sqrt{s}=(4681.92\pm0.30)~\mathrm{MeV}$.
No statistically significant signal is observed in the mass range from $4.08$ to $4.32~\gev/c^{2}$. The upper limits of 
$\sigma[\EE\to\gamma X]\cdot \mathcal{B}[X \to \ddstar]$ at a $90\%$ confidence level are determined.
\end{abstract}

\maketitle

\section{I. INTRODUCTION}
The charmonium(like) system is a good laboratory for studying the nonperturbative behavior of quantum chromodynamics (QCD). 
Over the past two decades, many new resonant structures have been discovered that cannot be explained by a simple mesonic~($q\bar{q}$) 
or baryonic~($qqq$) configurations. This has sparked significant experimental and theoretical interest.

Charmonium states with mass above the open-charm threshold are expected to decay dominantly into open-charm final states ~\cite{ozione,ozitwo,ozithree}. 
In Ref.~\cite{Higher_Charmonia},  the masses and total widths of the $\chi_{c2}(3P)$, $\chi_{c1}(3P)$, and $\eta_{c2}(2D)$ are 
calculated using either the nonrelativistic potential model or the relativized Godfrey-Isgur model~\cite{Higher_Charmonia1}.
The open-charm strong decay widths are estimated by using harmonic oscillator wave functions and the ${}^{3}{P}_{0}$ decay 
model. These states are predicted to have sizeable decay widths to the $\ddstar$ final state, as listed in Table~\ref{tab:X1}. 
However, none of these states have been observed experimentally so far. In an amplitude analysis of the $B^{+}\to J/\psi \phi K^{+}$ 
decay, a $1^{++}$ state was observed in the $\phi J/\psi$ final state~\cite{lhca,lhcb,lhcc},  with an average mass of  $4286^{+8}_{-9}~{\rm MeV}/c^{\rm 2}$, 
which is close to the predicted mass of the $\chi_{c1}(3P)$. Its assignment as a $\chi_{c1}(3P)$ state requires further confirmation. 

In the hybrid meson  ($q\bar{q} g$) configuration, several states are predicted by lattice QCD~\cite{QED1,QED2}. The strong 
decay widths of these hybrid states are computed based on the constituent gluon model~\cite{constituent_gluon}.
The $D_{\rm s}^{\pm}D_{\rm s}^{*\mp}$ mode is expected to be the leading decay channel for the $X(4217)$ with $J^{PC}=1^{-+}$ 
and the $X(4279)$ with $J^{PC}=0^{-+}$, where the numbers in the brackets denote the theoretical masses of the hybrid states. 
The resonant parameters of the two states are summarized in Table~\ref{tab:X1}. 

The authors of Ref.~\cite{heavy-antiheavy} provide a whole spectrum of heavy-antiheavy
hadronic molecules by solving the Bethe-Salpeter equation.~A virtual state, referred to as the $X(4080)$ in this paper, with a mass close to the $\ddstar$ threshold, is expected to decay predominantly to 
$\ddstar$. Its quantum numbers are $J^{PC}=1^{++}$ and its total width is assumed to be $5~\mev$.~Many structures from experiments with masses near the thresholds of a pair of open-charm mesons are reported, such as the $X(3872)$~\cite{x3872}, $Z_c(3900)^{\pm}$~\cite{zc3900a,zc3900b,zc3900c}, $Z_c(4020)^{\pm}$~\cite{zc4020a,zc4020b}, 
and $Z_{cs}(3985)$~\cite{zcs3985a,zcs3985b}, which are near the thresholds of $D^{0}\bar{D}^{*0}$, $D\bar{D}^{*}$, $D^{*}\bar{D}^{*}$, and 
$D_{s}^{(*)}\bar{D}^{(*)}$, respectively. Motivated by both experimental observations and theoretical calculations, searching for the structure near the threshold of $\ddstar$ is crucial.

\begin{table}[htbp]
\centering
\caption{Predicted masses ($M$), total widths ($\Gamma$), partial widths ($\Gamma_{\ddstar}$), and quantum numbers of the six $C$-even states.}
\resizebox{\linewidth}{!}{
\begin{tabular}{lcccc}\\
\hline\hline
State &   $M$~($\mathrm{MeV}/c^{2}$) & $\Gamma$~($\rm MeV$) & $\Gamma_{\ddstar}$~($\mathrm{MeV}$)& $J^{PC}$\\
\hline
$\eta_{c2}(2D)$~\cite{Higher_Charmonia} & $4158$ & $111$ & $18$ & $2^{-+}$ \\
$\chi_{c1}(3P)$~\cite{Higher_Charmonia}~ & $4271$ & $39$ & $9.7$ & $1^{++}$\\
$\chi_{c2}(3P)$~\cite{Higher_Charmonia}~ & $4317$ &  $66$ & $11$ & $2^{++}$\\
\hline
$X(4080)$~\cite{heavy-antiheavy}  & $4082.55$       & $5$ &- &$1^{++}$\\
$X(4217) $~{\cite{QED1,QED2,constituent_gluon}} &  $4217$  & $6$ & $6$ & $1^{-+}$\\
$X(4279)$~{\cite{QED1,QED2,constituent_gluon}} & $4279$   & $110$  & $34$ & $0^{-+}$\\
\hline\hline
\end{tabular}
}
\label{tab:X1}
\end{table}

It has been demonstrated that $C$-even states can be produced in $\EE$ collisions accompanied by photon emission.~For instance, 
the $\EE\to\gamma X(3872)$~\cite{gamma3872} and $\gamma\chi_{c1,2}(1P)$~\cite{gammachicj} processes have been observed 
by the BESIII experiment.~In this paper, a study of $\EE\to\gamma \ddstar$ is presented to search for the aforementioned $C$-even 
structures.~Throughout this paper, the charge conjugated mode is implied.~This analysis utilizes a data sample collected at center-of-mass 
energy~$\sqrt{s}=(4681.92\pm0.08\pm0.29)~\mathrm{MeV}$ with an integrated luminosity of $(1667.39\pm0.21\pm8.84)~\mathrm{pb}^{-1}$~\cite{BESIII:lum}, 
where the first and second uncertainties are statistical and systematic, respectively. The $D_{s}^{*}$ meson is reconstructed using 
its radiative decay into the $\gamma D_{s}$ final state. The $\ds$ meson is reconstructed with its decay modes $\ds \to \kk\pi$ and $\kshort K$, 
with $\kshort \to \pp$. These two decay modes of the $D_{s}$ are combined into four decay chains (DCs):\\
DC-I:~~~$\ds^{+}\ds^{-}\to (K^{+}K^{-}\pi^{+}) (K^{+}K^{-}\pi^{-})$,\\
DC-II:~~$\ds^{+}\ds^{-}\to(K^{+}K^{-}\pi^{+})(\kshort K^{-})$,\\
DC-III:~$\ds^{+}\ds^{-}\to(\kshort K^{+})(K^{+}K^{-}\pi^{-})$, and\\
DC-IV:~$\ds^{+}\ds^{-}\to(\kshort K^{+})(\kshort K^{-})$.

\section{II. BESIII DETECTOR AND MONTE CARLO SIMULATION}

The BESIII detector~\cite{bes} records symmetric $e^+e^-$ collisions 
provided by the BEPCII storage ring~\cite{Yu:IPAC2016-TUYA01}, which 
operates with a peak luminosity of $1\times10^{33}$~cm$^{-2}$s$^{-1}$
in the center-of-mass energy range from $2.0$~GeV to $4.95$~GeV~\cite{Ablikim:2019hff}.
The cylindrical core of the BESIII detector covers 93\% of the full 
solid angle and consists of a helium-based multilayer drift chamber 
(MDC), a plastic scintillator time-of-flight system (TOF), and a CsI(Tl)
electromagnetic calorimeter (EMC), which are all enclosed in a superconducting
solenoidal magnet providing a 1.0 T magnetic field.~The solenoid is supported by an octagonal 
flux-return yoke with resistive plate counter muon identification modules interleaved 
with steel. The charged-particle momentum resolution at $1~{\rm GeV}/c$ is $0.5\%$, 
and the ${\rm d}E/{\rm d}x$ resolution is $6\%$ for electrons from 
Bhabha scattering. The EMC measures photon energies with a
resolution of $2.5\%$ ($5\%$) at $1~\gev$ in the barrel (end cap)
region. The time resolution in the TOF barrel region is 68~ps, while
that in the end cap region is 60~ps~\cite{etofa,etofb,etofc}.

Simulated events produced with a {\textsc{geant4}}-based~\cite{geant4} Monte Carlo (MC) package are used to 
determine detection efficiencies and to estimate backgrounds.~This MC package includes the geometric description of 
the BESIII detector and the detector response. The simulation models the beam energy spread and initial state radiation 
(ISR) in the $\EE$ annihilations with the generator {\sc kkmc}~\cite{ref:kkmca,ref:kkmcb}. The generic MC samples include 
the production of open-charm processes, the ISR production of vector charmonium(like) states, and the continuum processes 
implemented in {\sc kkmc}~\cite{ref:kkmca,ref:kkmcb}.~The known decay modes are modeled with {\sc evtgen}~\cite{ref:evtgen,ref:evtgena} 
using branching fractions taken from the Particle Data Group~(PDG)~\cite{PDG2018}, and the remaining unknown 
charmonium decays are modelled with {\sc lundcharm}~\cite{ref:lundcharma,ref:lundcharmb}.  Final state radiation for 
charged final state particles is incorporated using the {\sc photos} package~\cite{ref:photos}.

The signal processes $\EE\to\gamma X,~X\to \ddstar$~and their subsequent decays are modeled with {\sc evtgen}~\cite{ref:evtgen,ref:evtgena}.
The $\ds\to\kk\pi$ decay is generated according to the amplitude models reported in Refs.~\cite{ref:da1,ref:da2}.
Two groups of samples are generated for the signal processes, Group-I and Group-II. Group-I is produced with the predicted masses 
and widths of the $X$ as listed in Table~\ref{tab:X1}, while Group-II is with the masses ranging from $4080~\mevcc$ to $4320~\mev/c^{2}$ in 
steps of $40~\mev/c^{2}$  and widths varying from a set of values: $5~\mev$, $10~\mev$, $20~\mev$, $50~\mev$, $80~\mev$, and $100~\mev$. The quantum 
numbers $J^{PC}$ of $X$ are listed in Table~\ref{tab:X1}. In the signal MC simulations of $e^{+}e^{-}\to \gamma X$, $\gamma X$ are assumed to generate from 
$\psi(4660)$, and $\psi(4660)$ is parametrized using a Briet-Wigner function with parameters taken from the PDG~\cite{PDG2018}.

Additional MC samples of $\EE$$\to$$(\gamma_{\rm ISR})D_{s}^{(*)\pm}D_{s}^{*\mp}$, $(\gamma_{\rm ISR})D_{s}^{*}D_{s0}^{*}(2317)$, 
and $(\gamma_{\rm ISR}) D_{s}^{(*)} D_{s1}(2460)$~events are generated to estimate the background contamination. They are modeled 
using helicity-amplitude models or phase space in {\sc evtgen}~\cite{ref:evtgen,ref:evtgena}.~The production cross sections are taken from 
previous BESIII measurements~\cite{ref:dsstardsstar,ref:dsds2460,ref:dssds2460}.

\section{III. EVENT SELECTION AND BACKGROUND STUDY}
A full reconstruction method is used to reconstruct the signal process $\gamma\ddstar$($\to\gamma\ds^{\mp}$), $\ds\to\kk\pi$ or 
$\kshort(\to\pi^{+}\pi^{-}) K$. Charged tracks detected in the MDC are required to be within a polar angle ($\theta$) range of 
$|\cos\theta|<0.93$, where $\theta$ is defined with respect to the symmetry axis of the MDC (defined as the $z$-axis). For charged 
tracks not originating from $\kshort$ decays, the distance of the closest approach to the interaction
point (IP) must be less than $10~\mathrm{cm}$ along the $z$-axis ($V_{z}$), and less than $1~\mathrm{cm}$ in the transverse plane 
($V_{xy}$). Particle identification (PID) for charged tracks combines measurements of the energy deposited (${\rm d}E/{\rm d}x$) in the 
MDC  and the flight time in the TOF to form a likelihood for each hadron hypothesis.~Tracks are identified as kaons (pions) when the kaon 
(pion) hypothesis has larger probability. Tracks without PID information are rejected.

Each $\kshort$ candidate is reconstructed from two oppositely charged tracks satisfying $|V_z|<20~\mathrm{cm}$ and 
$|\cos\theta|<0.93$. The two charged tracks are assigned as $\pi^+ \pi^-$ without imposing PID criteria. They are constrained 
to originate from a common vertex and are required to have an invariant mass $(M_{\pi^+\pi^-})$ within the interval $(0.478,~0.518)$~GeV/$c^2$. 
The $\chi^{2}$ of the vertex fit is required to be less than $100$. A secondary vertex fit is performed to ensure that the $\kshort$ 
momentum points back to the IP. The decay length of the $\kshort$ candidate is required to be greater than twice the vertex resolution.

Photon candidates are identified using isolated showers not associated with charged tracks in the EMC. The deposited energy 
of each shower must be greater than $25~\mev$ in the barrel region 
($|\cos\theta|<0.8$) and greater than $50~\mev$ in the end cap regions 
($0.86<|\cos\theta|<0.92$). The difference between the EMC time and the 
event collision time is required to be within $[0,~700]~\mathrm{ns}$ to 
suppress electronic noise and showers unrelated to the event.

The selected $K^{\pm}$, $\pi^{\pm}$, and $\kshort$ candidates in the event are combined to reconstruct $D_{s}\to K^{+}K^{-}\pi$ 
or \mbox{$D_{s}\to\kshort K$.} 
The $D_s$ candidates are kept if the invariant masses of the $\kk\pi$ or $\kshort K$ systems, denoted as $M_{D_{s}}$,
are within a $\pm 150~\mevcc$ mass window around the nominal $D_s$ mass. 
The $\ds^{+}\ds^{-}$ pairs are selected requiring that each $K^{\pm}$, $\pi^{\pm}$, and $\kshort$ candidate is used in one 
$\ds^{+}\ds^{-}$ pair at most once. If there is more than one pair with the same DC in an event, the one with $(M_{\ds^{+}}+M_{\ds^{-}})/2$ 
closest to the nominal mass of the $\ds$ meson~\cite{PDG2018} is chosen.

In addition to the $\ds^{+}\ds^{-}$ pair, each signal candidate is required to have at least two photons.~A six-constraint (6C) kinematic fit 
is performed, where the four-momentum of the final state particles is constrained to that of the initial $\EE$ system and the masses of 
the reconstructed $D_s^{+}$ and $D_{s}^{-}$ are constrained to their nominal masses. 
If more than two photons and (or) more than one $\ds^{+}\ds^{-}$ pair decaying in different DCs are found in an event, the combination 
with the minimum $\chi^2_{\rm 6C}$ is retained, and $\chi^2_{\rm 6C}$ is further required to be less than 200, where $\chi^2_{\rm 6C}$ 
is the $\chi^{2}$ from the 6C kinematic fit. To study the background contributions from non-$D_{s}$ processes, a four-constraint kinematic 
(4C) fit without the two $D_s$ mass constraints is applied, and the four-momenta of the final state particles are computed using the momenta 
determined by the 4C kinematic fit.

To suppress background contributions from non-$D_{s}$ processes, a $D_{s}$ mass window of $|M_{D_{s}}-m_{D_{s}}|<21~\mev/c^{2}$, 
which is about three times the mass resolution, is applied, where $m_{D_{s}}$ is the nominal $D_{s}$ mass. 
The two selected photon candidates and the two $D_{s}$ mesons can form four combinations of $\gamma\ds$. The one with invariant mass 
($M_{ \gamma\ds})$ closest to the nominal mass of the $D_s^{*}$ is retained for further analysis. 
The invariant mass $M'_{\gamma\ds}=(M_{\gamma\ds}-M_{\ds}+m_{\ds})$ is required to be within the interval 
$(m_{D_{s}^{*}}\pm27)~\mev/c^{2}$, where $m_{D_{s}^{*}}$ is the nominal $D_{s}^{*\pm}$ mass taken from the PDG~\cite{PDG2018}.
The variable $M_{\gamma\ds}$ formed with the unused photon and $\ds$ candidates
is required to be outside of this same mass window. The photon not from the $\dsstar$ decay is 
labeled $\gamma_{1}$ and its recoil mass spectrum, $M_{\gamma_{1}}^{\rm rec}$, 
is used to extract the signal yields. $M_{\gamma_{1}}^{\rm rec}$ is calculated by $\sqrt{(E_{\rm cms}-E_{\gamma_{1}})^{2}-(\vec{P}_{\rm cms}-\vec{P}_{\gamma_{1}})^{2}}$, 
where ($E_{\rm cms}, \vec{P}_{\rm cms}$) and ($E_{\gamma_{1}}, \vec{P}_{\gamma_{1}}$) are the four-momenta of 
the initial state~\cite{BESIII:lum} and $\gamma_{1}$, respectively.~After imposing all the selection requirements, the dominant 
background processes are those that include one $D_{s}^{+}D_{s}^{-}$ pair in final states.
The contributions from  $\EE$$\to$$(\gamma_{\rm ISR})D_{s}^{(*)\pm}D_{s}^{*\mp}$, $(\gamma_{\rm ISR})D_{s}^{*} D_{s0}^{*}(2317)$, 
and $(\gamma_{\rm ISR}) D_{s}^{(*)} D_{s1}(2460)$ events account for $76.9\%$, $4.4\%$, and $10.9\%$ of the total backgrounds, 
respectively.~The contributions from other background processes are less than $8\%$, and are found to be linearly distributed in the recoil 
mass spectrum of $\gamma_{1}$ based on the analysis of generic MC samples. 

\section{IV. SIGNAL YIELD EXTRACTION}

The $X$ signal yields are extracted using an unbinned maximum likelihood fit to the $M_{\gamma_{1}}^{\rm rec}$
spectrum.~The events from the four DCs are combined due to limited statistics. The fit probability density function
(PDF) contains five components: signal ($S$), $\EE\to(\gamma_{\rm ISR})\ddstar$ background ($B_{1}$), 
\mbox{$\EE\to(\gamma_{\rm ISR})\dstardstar$} background ($B_{2}$), $\EE\to(\gamma_{\rm ISR})D_{s}^{(*)}D_{sJ}$ 
background ($B_{3}$), and other remaining backgrounds ($B_{4}$). The function used in the fit is defined as
\begin{linenomath*}
\begin{equation}
\begin{aligned}
 f_{\rm sum} = N_{\rm sig} \cdot S + \sum_{i=1}^{4}N_{\mathrm{bkg} i}\cdot B_{i} ,\nonumber
\end{aligned}
\end{equation}
\end{linenomath*}
where $N_{\rm sig}$~and~$N_{\mathrm{bkg}(1,2,3,4)}$ represent the numbers of signal events and 
background events, respectively. $N_{\mathrm{bkg}(1,2,4)}$ are free parameters, and $N_{\mathrm{bkg}3}$ 
is fixed according to BESIII measurement in the fit, $N_{\mathrm{bkg}3}=8.8\pm 0.9$~\cite{ref:dsds2460,ref:dssds2460}, 
where the number of events and the uncertainty are from the cross section measurements. 

The signal PDF is described by $S=BW(M_{\gamma_1}^{\rm rec};M,\Gamma)\otimes DG(m_{1},\sigma_{1},m_{2},\sigma_{2},f)\otimes G(\Delta m,\Delta \sigma)$, 
where $BW$ stands for the Breit-Wigner function with mass $M$ and width $\Gamma$ fixed, $DG$ is a double Gaussian function that describes the 
mass resolution and mass shift in the simulation, and $G$ is a Gaussian function that 
accounts for the mass resolution difference between data and MC simulation.~The parameters $m_{1}(m_{2})$ and $\sigma_{1} (\sigma_{2})$ represent 
the mean and standard deviation of the first (second) Gaussian function in the double Gaussian model, while $f$ is the fraction of the 
first Gaussian component over the whole $DG$. To get the parameters of the double Gaussian function, the $M_{\gamma_1}^{\rm rec}-M_{\gamma_1}^{\rm truth}$ 
distribution from the signal MC samples is fitted, where $M_{\gamma_1}^{\rm truth}$ is the recoil mass spectrum of the $\gamma_1$ distribution
at generator level. The parameters of the Gaussian function are obtained by using a control sample, $\EE\to\ddstar$.~The selection of the control 
sample is similar to that of the signal sample, except that only one photon is required instead of two. The $D_{s}$ and $D_{s}^{*}$ mass distributions from 
data are fitted with the line shape from MC samples convoluted with Gaussian functions.~The parameters of the Gaussian function, representing 
the difference of the mass resolutions between data and MC simulation, are then used as the parameters of the Gaussian function in the extraction 
of the $X$ signal yield.~The $B_{1(2)}$ line shape is taken from the $\EE\to(\gamma_{\rm ISR})\ddstar$ ($\EE\to(\gamma_{\rm ISR})\dstardstar$) MC simulation,
$B_{3}$ is from the sum of the $\EE\to(\gamma_{\rm ISR})D_{s}^{*}D_{s0}^{*}(2317)$ and $(\gamma_{\rm ISR}) D_{s}^{(*)} D_{s1}(2460)$ MC 
simulations (weighted according to the product of the cross sections and detection efficiencies), and $B_{4}$ is modeled by a polynomial function.

\begin{figure*}[htp]
  \centering
\includegraphics[scale=0.4]{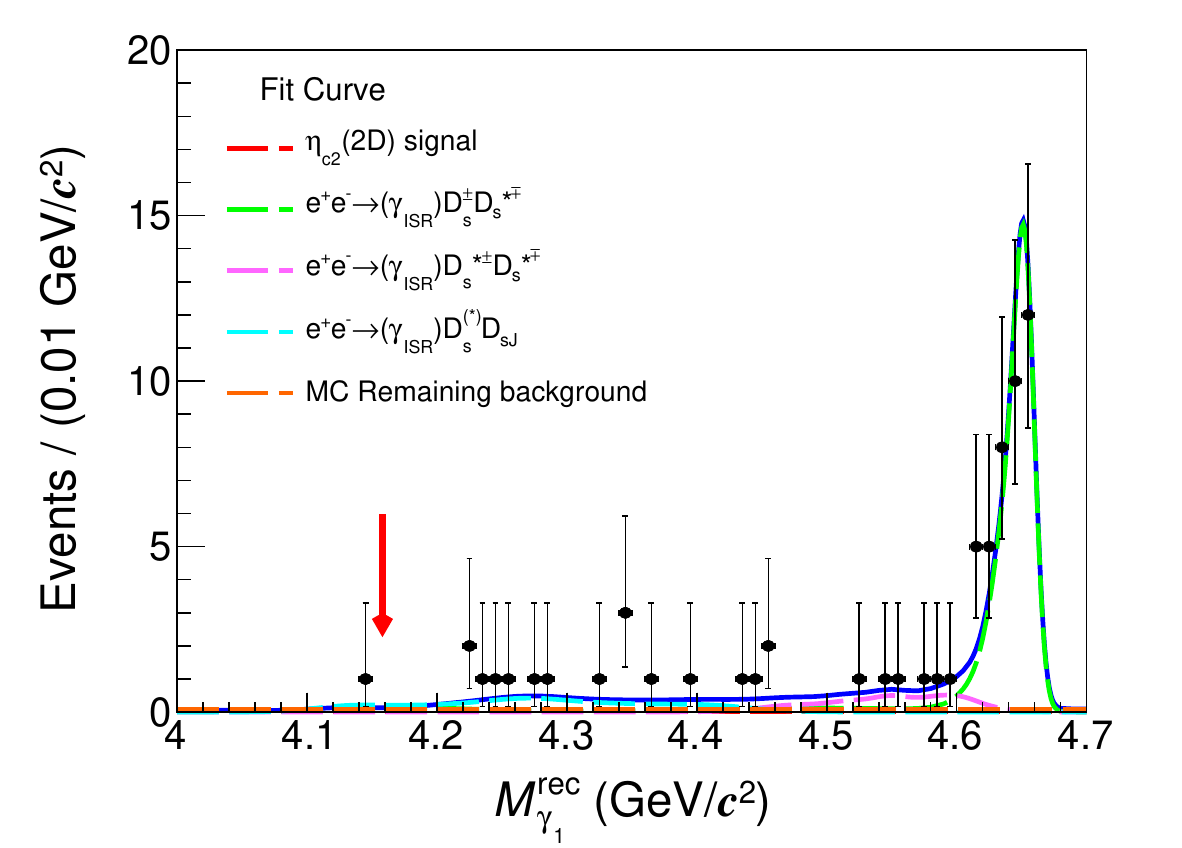}
   \includegraphics[scale=0.4]{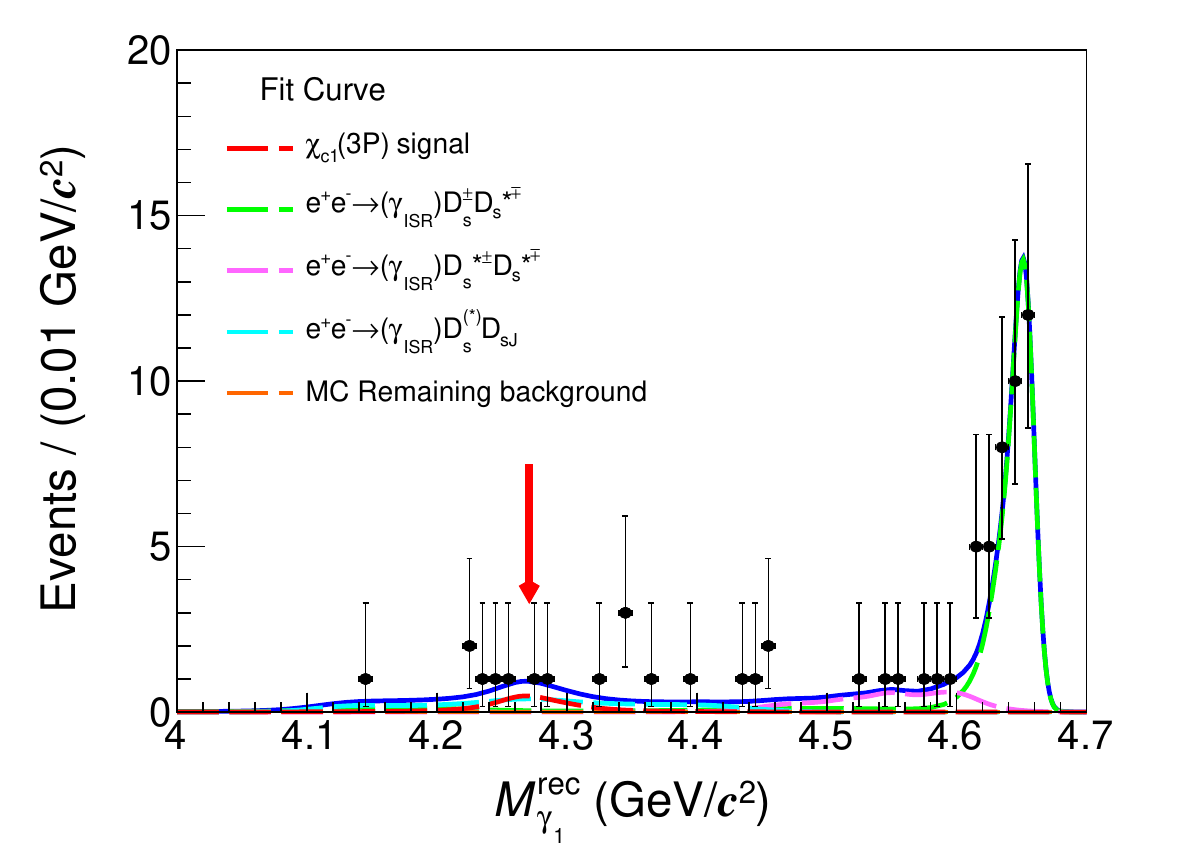}
    \includegraphics[scale=0.4]{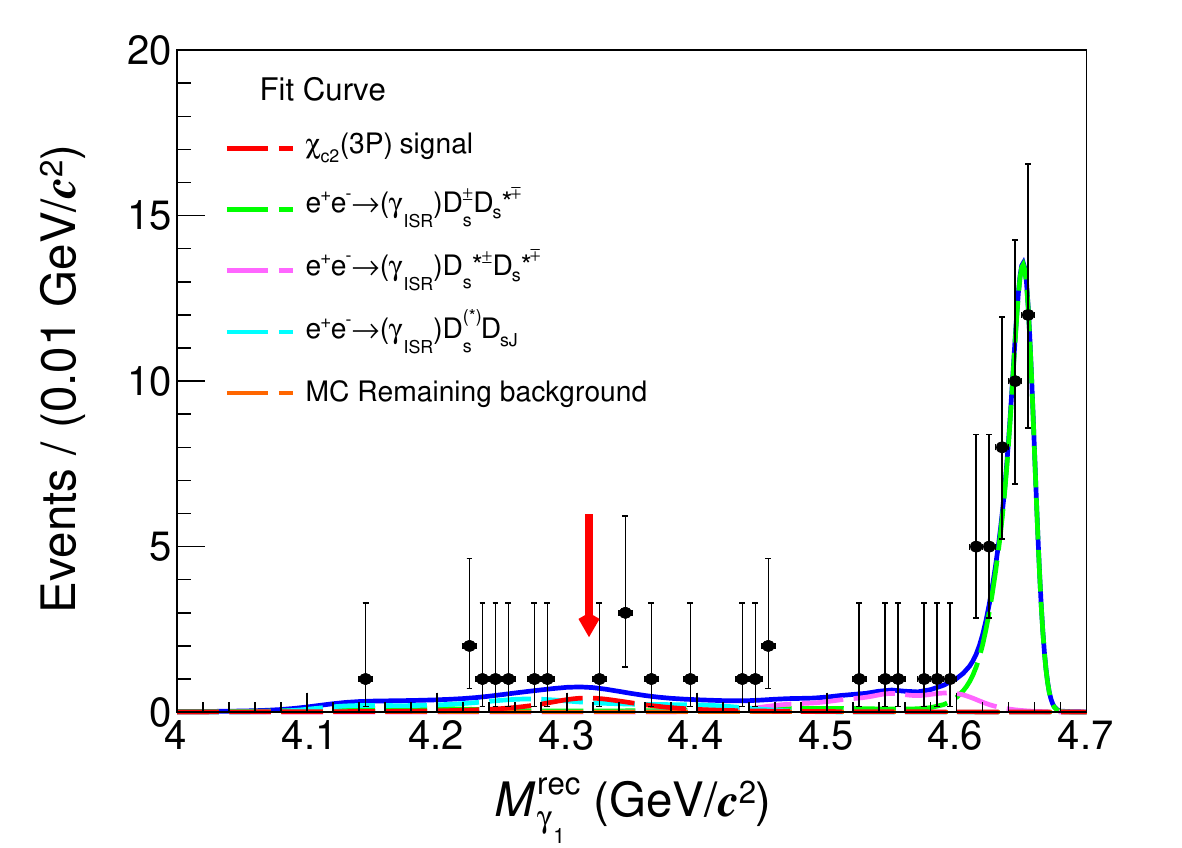}
   \includegraphics[scale=0.4]{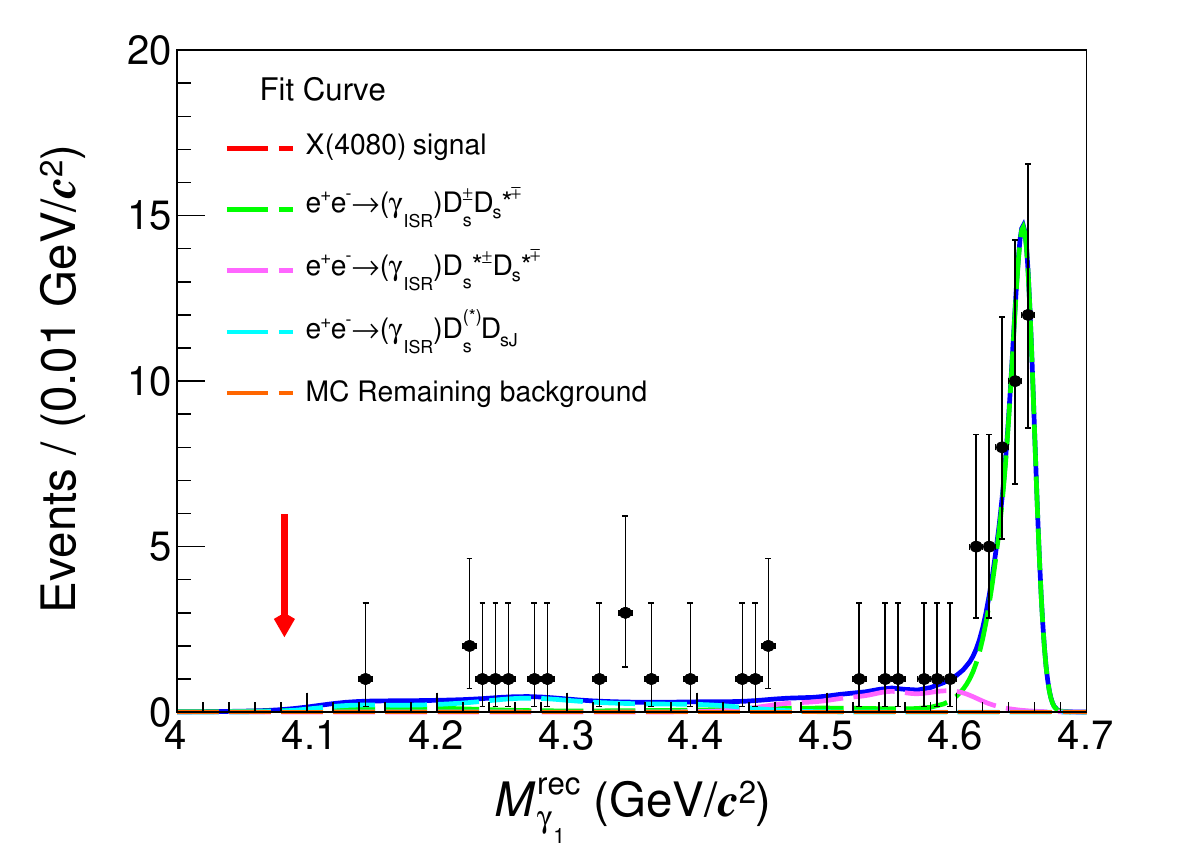}
   \includegraphics[scale=0.4]{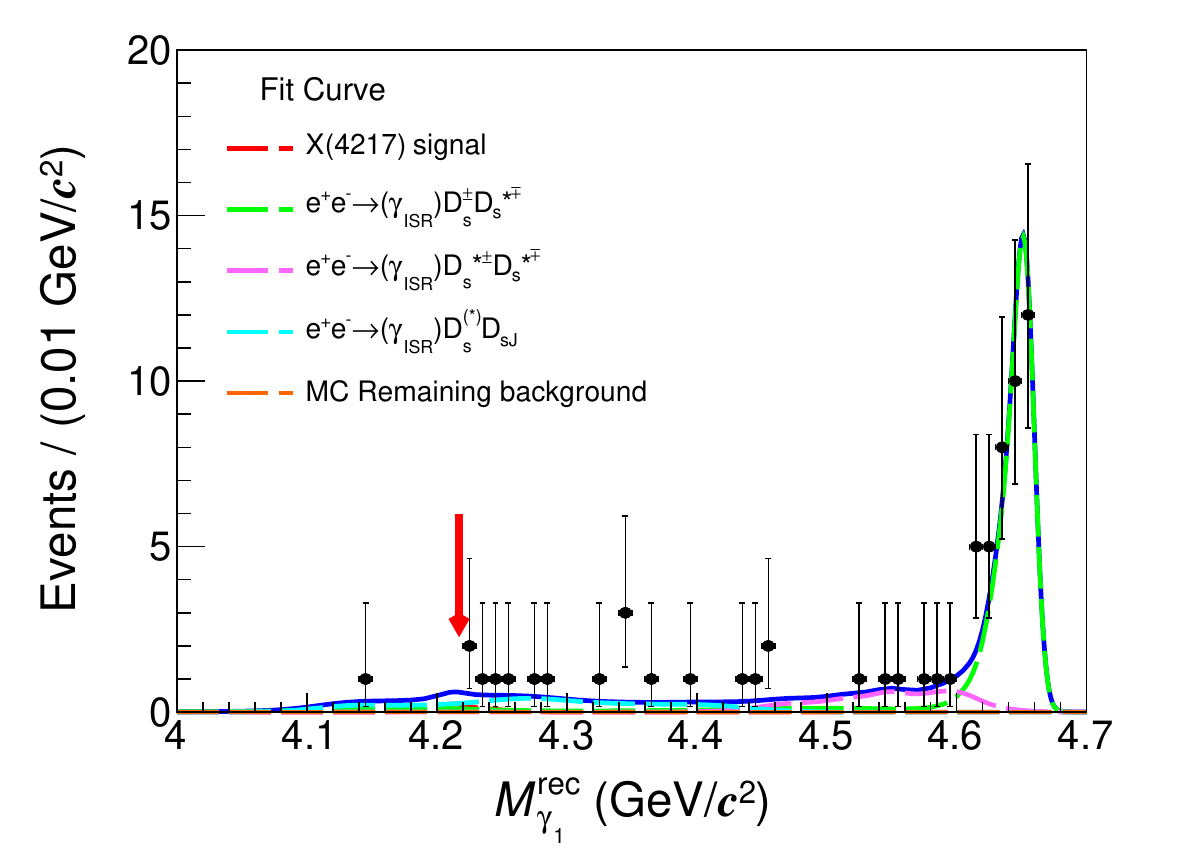}
    \includegraphics[scale=0.4]{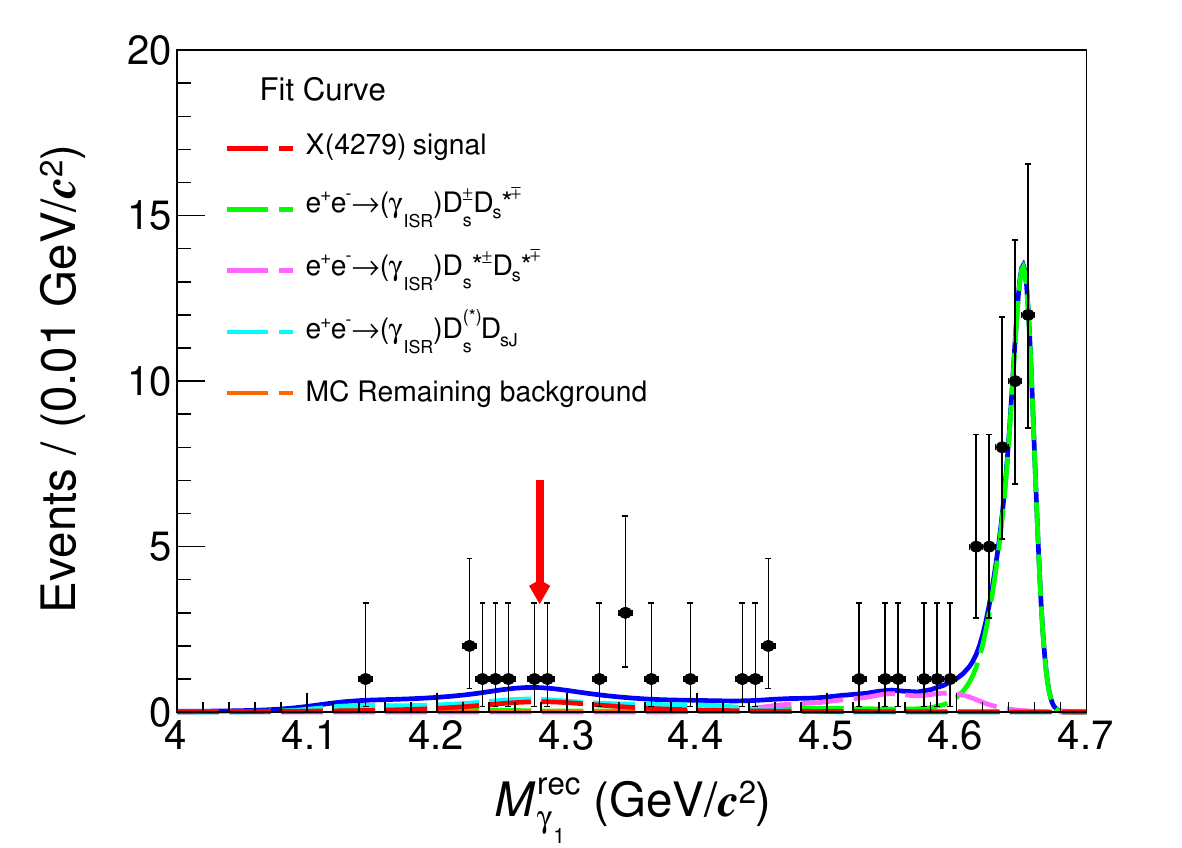}
  \caption{Fits to the $M_{\gamma_{1}}^{\rm rec}$ spectrum with masses and 
  widths fixed to those of the $\eta_{c2}(2D)$,
  $\chi_{c1}(3P)$, $\chi_{c2}(3P)$, $X(4080)$, $X(4217)$, and $X(4279)$ as listed in Table~\ref{tab:X1}. The 
  dots with error bars are the data distribution and the blue solid lines are 
  the best fit results. The red arrows indicate the peak positions of those six $X$ states with masses fixed to the predictions listed in Table~\ref{tab:X1}.} 
  \label{fig:meanexcitefit}
\end{figure*}

Figure~\ref{fig:meanexcitefit} shows the fit results where the masses and widths of the $X$ states are fixed to the predictions listed 
in Table~\ref{tab:X1}. The signal yields are summarized in Table~\ref{tab:6}.~Since no events are retained near the $\ddstar$ 
mass threshold, the number of $X(4080)$ signal yields is calculated using $N_{\rm obs} - N_{\mathrm{bkg}(1,2,3,4)}$, where 
$N_{\rm obs} = 0$ is the number of observed events, $N_{\mathrm{bkg}(1,2,3,4)}$ are the numbers of background events obtained 
from the fit with $N_{\rm sig}$ set to zero. All the numbers are counted within $[4.068,~4.098]~{\rm GeV}/c^{2}$, which is about 
$3\sigma$ around the $X(4080)$ nominal mass.
The statistical significance is evaluated by using $\sqrt{2\ln(L_{\rm max}/L_{\rm 0})]/(\Delta\mathrm{n.d.f}=1)}$, where 
$\ln L_{\rm max}$ and $\ln L_{0}$ are the logarithmic likelihood values with and without the signal component in the 
fits, and $\Delta\mathrm{n.d.f}$ is the change of the number of degrees of freedom.
The upper limits of the number of signal events are calcualted using a Bayesian method~\cite{ref:Bayesianmethod} with a  prior.
The likelihood distribution, $L(x)$, is obtained by repeating the fit to the $M_{\gamma_{1}}^{\rm rec}$ spectrum with a series 
of fixed values for the number of signal events, $x$. The upper limits at 90\% confidence level (C.L.) are determined from 
$\int_{0}^{N_{\rm sig}^{\rm UL}} L(x)dx/\int_{0}^{\infty}L(x)dx=0.9$, as listed in Table~\ref{tab:6}. 

\begin{figure}[htp]
  \centering
  \includegraphics[scale=0.35]{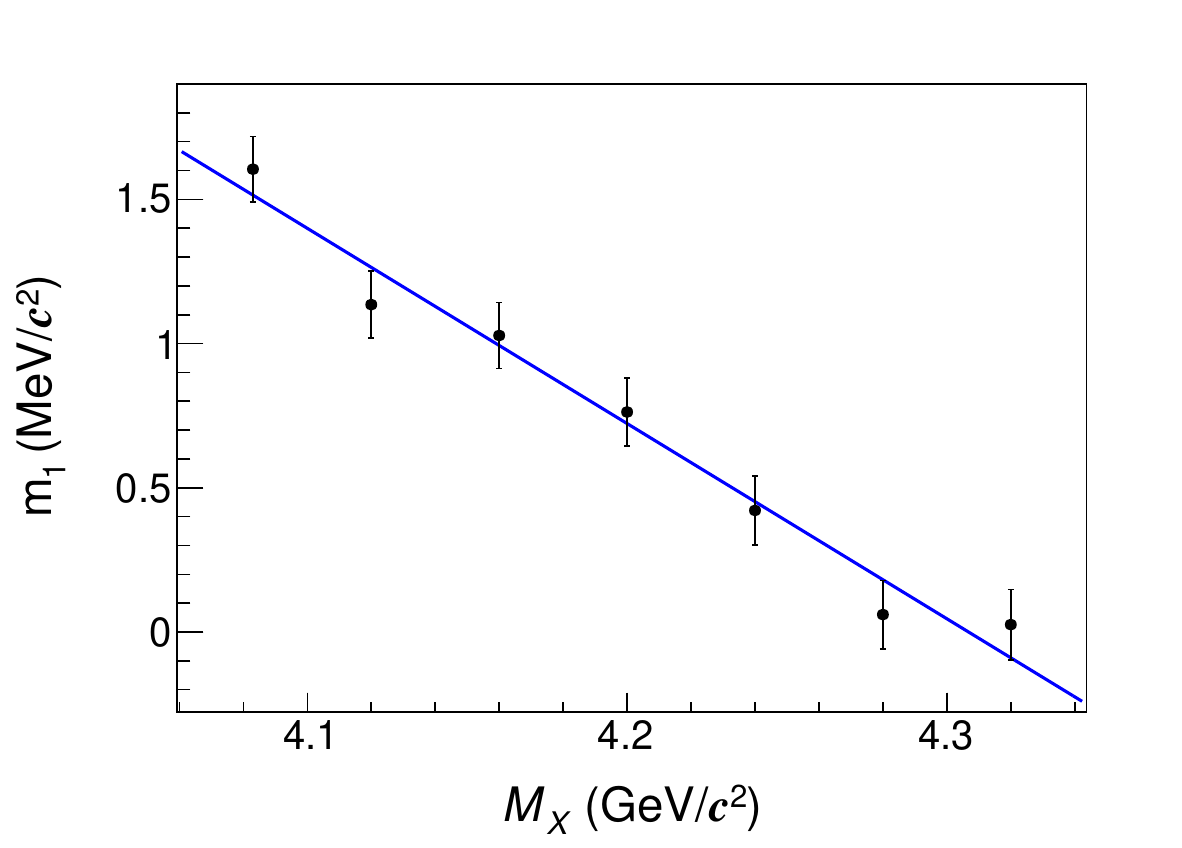}
   \includegraphics[scale=0.35]{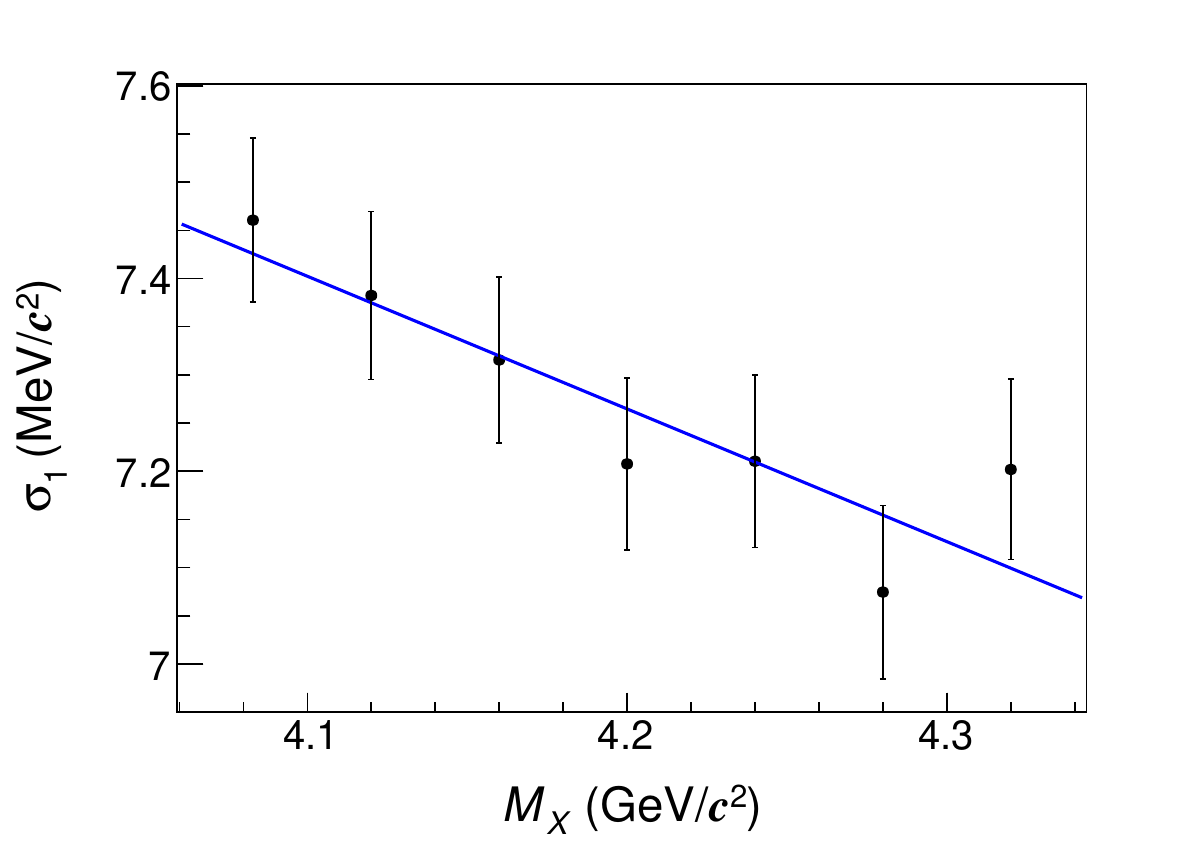}
  \caption{$M_{\rm X}$ dependence of $m_{1}$ (upper) and $\sigma_{1}$ (lower) for $\Gamma=50~\mev$  and 
  $J^{PC}$ = $0^{-+}$ according to MC simulation studies. The solid lines are the fit results.}
  \label{fig:m-sigma}
\end{figure}

\begin{figure*}[htp]
 \includegraphics[width=0.99\textwidth]{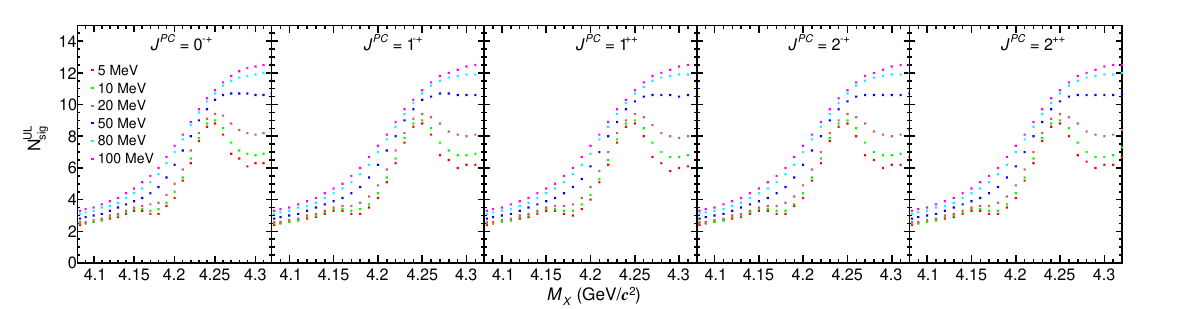}
  \caption{Upper limits on the number of $X$ signal events as a function of $X$ mass 
  for different quantum numbers ($J^{PC}=0^{-+}$, $1^{-+}$, $1^{++}$, $2^{-+}$, and $2^{++}$) and widths of the 
  $X$ ($\Gamma=5~\mev$, $10~\mev$, $20~\mev$, $50~\mev$, $80~\mev$, and $100~\mev$).}
  \label{fig:extuppfit}
\end{figure*}

In addition, upper limits for the signal cross sections as a function of the $X$ masses in the range $[4.08,~4.32]~\gevcc$ are 
also provided. From studies of simulation samples (Group-II), it is found the parameters 
$m_{2}-m_{1}$, $\sigma_{2}/\sigma_{1}$, and $f$ of the resolution function are very similar for various widths and $J^{PC}$.~The parameters are found to be
\begin{linenomath*}
\begin{equation}
\begin{aligned}\label{eq:einstein}
m_2 - m_1  & = (15.6 \pm 0.2) ~\rm{MeV/c^{2}}, \\
\sigma_2/\sigma_1 & = 3.01 \pm 0.02, \\
f & = 0.692 \pm 0.005,
\end{aligned}
\end{equation}
\end{linenomath*} 
using the MC sample with $\Gamma=50~\mev$ and $J^{PC}=0^{-+}$, and they are fixed in the latter fit to $M_{\gamma_{1}}^{\rm rec}$.
The $m_{1}$ and $\sigma_{1}$ dependence on the mass of the $X$ state are shown in Fig.~\ref{fig:m-sigma}. A parametrization of the 
resolution, as a function of $m_{1}$ and $\sigma_{1}$ vs. $M_{\rm X}$ distributions, is fitted with first order polynomial functions. 
The dependencies of $m_{1}$ and $\sigma_{1}$  on other width and $J^{PC}$ are studied separately. The trends are similar to those 
shown in Fig.~\ref{fig:m-sigma}. The mass resolutions of the $X$ in the region $[4.08,~4.32]~{\rm GeV}/c^{2}$ are evaluated from 
these fitted function. The upper limits of the number of signal events as a function of the $X$ mass for different $X$ widths and quantum 
numbers are shown in Fig.~\ref{fig:extuppfit}.

\section{V. CROSS SECTION CALCULATION}
The Born cross section of $\EE\to\gamma X$ times the branching fraction of $X\to\ddstar$ is calculated as 
\begin{linenomath*}
\begin{align}
    \sigma \cdot \mathcal{B}[X\to\ddstar] 
    =\frac{N_{\rm sig}} {\mathcal{L}_{\rm int} f^{r} f^{v} \mathcal{B}[D_{s}^{*}\to\gamma D_{s}]\bar{\epsilon}},
\end{align}
\end{linenomath*}
where $N_{\rm sig}$ is the number of signal events obtained from the fit;~$\bar{\epsilon}=\sum_{j,k=1}^{2}\epsilon_{jk}\mathcal{B}_{j}\mathcal{B}_{k}$ 
represents the average detection efficiency, where $\epsilon_{jk}$ is the detection efficiency for
\mbox{$\EE\to\gamma X, X\to\ddstar, D_{s}^{*\mp} \to \gamma \ds^{\mp},$} 
\makebox{$\ds^{+}\to j, \ds^{-}\to k$; and $\mathcal{B}_{j}$
($\mathcal{B}_{k}$)} is the branching fraction of $\ds^{+}\to j$ ($D_{s}^{-}\to k$). Here $j$ and $k$ are the first and second decay modes of the $D_{s}$.
The ISR correction factor ($f^{r}=(1+\delta)$) is calculated following the procedure described in Refs.~\cite{ref:kkmca,ref:kkmcb}. The vacuum polarization 
correction factor is $f^{v}=1/|1-\Pi|^2$~\cite{Jegerlehner:2011mw}.~$\mathcal{B}[{D_{s}^{*}\to\gamma D_{s}]}$ is the branching fraction of 
$\dsstar\to\gamma\ds$. The results are summarized in Table~\ref{tab:6}. The product of the upper limit of the cross section and the branching fraction 
ranges between $4$~pb and $35$~pb, depending on the mass and width of the $X$ states.

\begin{table*}[htp]
 \centering
\caption{Product of upper limits of Born cross section of $\EE\to\gamma X$ and the branching fraction of $X\to\ddstar$ at 90\% C.L. for 
each candidate $C$-even state, where $f^{r}$, $f^{v}$, $N_{\rm sig}$, $N_{\rm sig}^{\rm UL}$, $N_{\mathrm{bkg}1,2,4}$, 
 and $\bar{\epsilon}$ are the ISR correction factor, the vacuum polarization correction factor, the number of signal events, the upper limit on the number of 
 signal events, the numbers of background events, and the average detection efficiency, respectively. ``$\sigma^{\mathrm{UL}} \cdot \mathcal{B}$ with sys." 
 stands for the upper limits of the cross section times the branching ratio with systematic uncertainties.~``Significance" represents statistical significance.}
 \begin{tabular}{c|ccccccc}
  \hline \hline
   & $\eta_{c2}(2D)$ & $\chi_{c1}(3P)$ & $\chi_{c2}(3P)$ & $X(4080)$ & $X(4217)$ &  $X(4279)$ \\
  \hline
   $f^r$& $1.06$ & $1.06$ & $1.06$  & $1.06$ &  $1.06$ & $1.06$\\
 $f^v$& $1.05$ & $1.05$ & $1.05$ & $1.05$ & $1.05$ & $1.05$\\
   $N_{\rm sig}^{\rm UL} $ & $5.2$ &$10.1$ & $11.4$& $2.4$ &  $6.0$ &  $12.2$ \\
  $N_{\rm sig} $ & $-4.6^{+3.6}_{-2.9}$ & $4.0^{+ 3.8}_{- 3.0}$ & $5.0^{+ 4.1}_{- 3.3}$ & $-0.3^{+0.3}_{-0.0}$ & $1.1^{+ 2.7}_{- 1.9}$ & $5.3^{+ 4.5}_{- 3.7}$ \\
   $N_{\rm bkg1} $ & $43.1^{+ 7.6}_{- 7.0}$ & $42.3^{+ 7.4}_{- 6.7}$ &  $42.2^{+ 7.4}_{- 6.7}$ & $42.9^{+7.5}_{-6.8}$ &  $42.7^{+ 7.5}_{- 6.8}$ & $42.1^{+ 7.4}_{- 6.7}$ \\
    $N_{\rm bkg2} $ & $6.6^{+ 4.8}_{- 4.1}$ & $8.2^{+ 4.3}_{- 3.4}$ & $7.9^{+ 4.2}_{- 3.4}$& $8.2^{+4.3}_{-3.8}$ &  $8.3^{+ 4.3}_{- 3.4}$ & $7.8^{+ 4.2}_{- 3.4}$ \\
    $N_{\rm bkg4} $ & $6.0^{+ 9.3}_{- 0.0}$ & $0.0^{+ 2.9}_{- 0.0}$ & $0.0^{+ 2.3}_{- 0.0}$& $0.0^{+ 6.0}_{- 0.0}$ &  $0.0^{+ 5.2}_{- 0.0}$ & $0.0^{+ 2.3}_{- 0.0}$ \\
  Significance $(\sigma)$ & $-$& $1.4$ & $1.6$& $-$ &  $0.5$ &  $1.5$  \\
  $\bar{\epsilon}~(10^{-4})$& $3.73$ &$3.48$ & $3.26$ & $4.21$ &  $3.50$ &  $3.11$ \\\hline
  $\sigma^{\rm UL} \cdot \mathcal{B}$ with sys. (pb) & $10.7$ & $23.4$ & $28.3$& $4.1$ &  $13.5$ & $32.4$ \\
 
  \hline\hline
 \end{tabular}
\label{tab:6}
\end{table*}

\begin{figure*}[htp]
  \centering
  \includegraphics[width=0.99\textwidth]{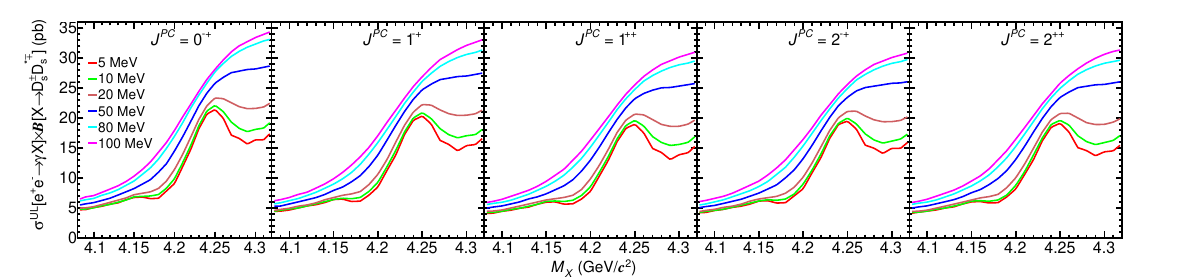}
  \caption{Upper limits on $\sigma[\EE\to\gamma X]\cdot \mathcal{B}[X\to\ddstar]$ as
  a function of $X$ mass for different quantum numbers ($J^{PC}=0^{-+}$, $1^{-+}$, $1^{++}$, $2^{-+}$, and $2^{++}$) and widths of the 
  $X$ ($\Gamma=5~\mev$, $10~\mev$, $20~\mev$, $50~\mev$, $80~\mev$, and $100~\mev$) after 
  considering the systematic uncertainties.}
  \label{fig:newborn}
\end{figure*}

\begin{figure*}[htp]
  \centering
 \includegraphics[width=0.9\textwidth]{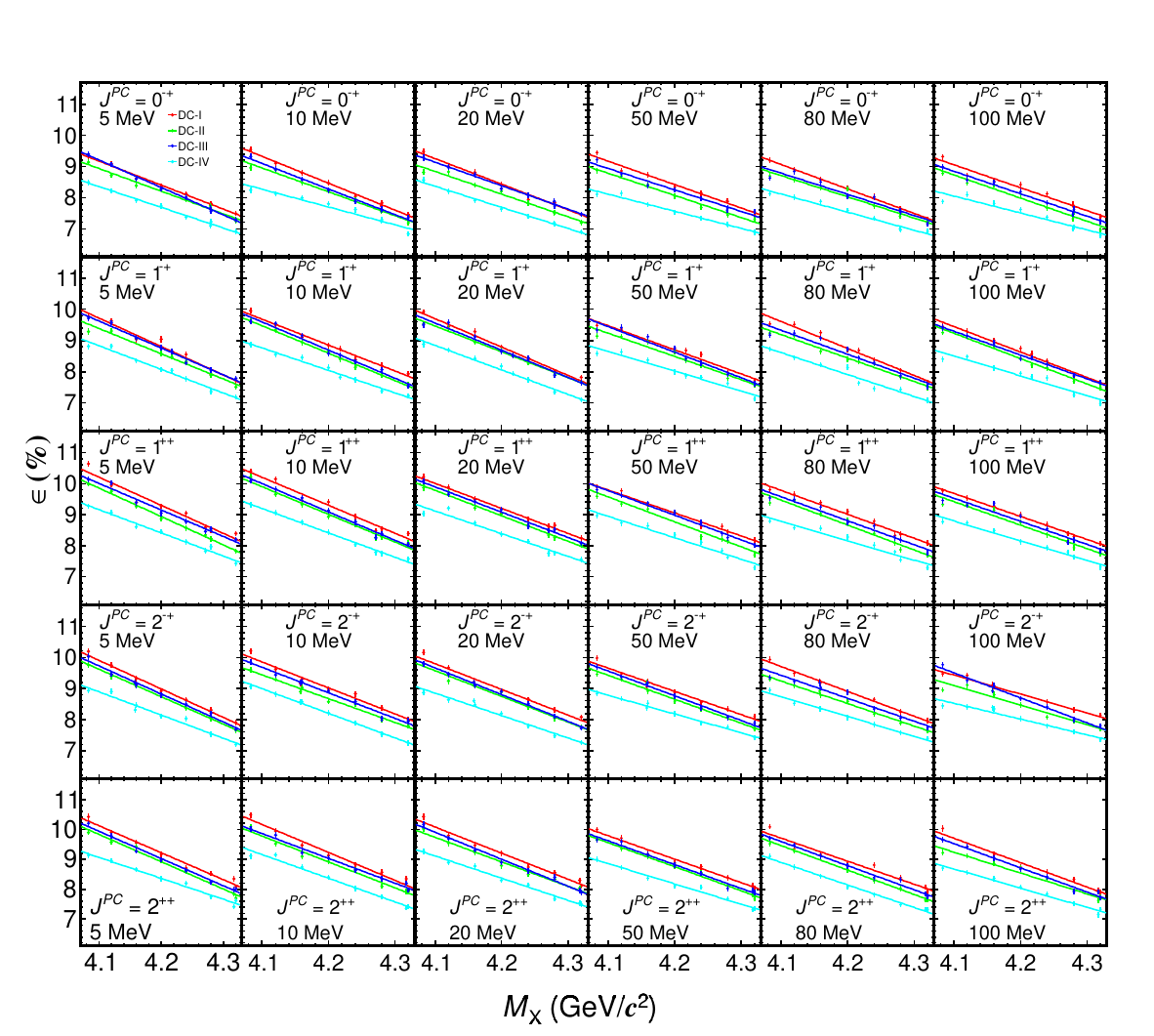}
  \caption{$M_{\rm X}$ dependence of the efficiencies for the four channels for different $J^{PC}$ and width assumptions.~The lines are the fit results.}
  \label{fig:effextfit}
\end{figure*}

Figure~\ref{fig:newborn} illustrates the upper limits of \mbox{$\sigma^{\rm UL} \cdot \mathcal{B}[X\to\ddstar]$} under different assumptions of the mass, width,
and $J^{PC}$ of the $X$ states after considering all the systematic uncertainties.~Detailed descriptions of the systematic uncertainties are provided 
in the subsequent sections. The detection efficiency as a function of $M$ for different width and $J^{PC}$ assumptions of the $X$ state is estimated from 
Group-II MC simulations. The $M_X$ dependence of the detection efficiency is fitted with a first order polynomial function, as shown in Fig.~\ref{fig:effextfit}.


In the calculation of the upper limits of Born cross sections displayed in Fig.~\ref{fig:newborn}, the detection efficiencies are extracted from the fitted curves 
and combined into the average detection efficiency $\bar{\epsilon}$.

\section{VII. SYSTEMATIC UNCERTAINTY ESTIMATION}

The systematic uncertainties on $\sigma \cdot \mathcal{B}$ are mainly originated from the luminosity measurement, the detection
efficiency, the quoted branching fractions, the ISR correction factor, and the fit models. The uncertainty from the vacuum polarization 
correction is negligible. A summary of systematic uncertainties is given in Table~\ref{tab:7}. The total systematic uncertainty is taken 
as a quadrature sum of each contribution.

\begin{table*}[htp]
\centering
\caption{Summary of systematic uncertainties (in $\%$) for the Born cross section measurements.}
\begin{tabular}{c|ccccc} 
  \hline \hline
   \makebox[0.08\textwidth]{Source}                  & \makebox[0.1\textwidth]{$J^{PC}=0^{-+}$} & \makebox[0.08\textwidth]{$1^{-+}$} & \makebox[0.08\textwidth]{$1^{++}$} & \makebox[0.08\textwidth]{$2^{-+}$} & \makebox[0.08\textwidth]{$2^{++}$}\\\hline
    Integrated luminosity                  & $1.0$      & $1.0$   & $1.0$      & $1.0$      & $1.0$\\
    PID                                   & $5.7$      & $5.7$     & $5.7$       &$5.7$      & $5.7$\\ 
    Tracking                                & $5.7$      & $5.7$     & $5.7$       &$5.7$      & $5.7$\\ 
    Photon reconstruction                 & $2.0$      &  $2.0$     & $2.0$       &$2.0$      & $2.0$\\ 
    $K_{S}^{0}$ reconstruction                &  $0.4$      &$0.4$      & $0.4$       &$0.4$       &$0.4$\\ 
    Kinematic fit                              &$0.7$        &$0.7$       &$0.8$         &$0.8$       &$0.7$\\ 
    $D_s$ mass window                  &$0.3$       &$0.3$        &$0.3$         &$0.3$       &$0.3$\\ 
    $D_s^*$ mass window                    &$0.1$               &$0.1$              &$0.1$                 &$0.1$              &$0.1$\\ 
    ISR correction factor               &$24.6$     &$24.4$   &$23.8$   &$23.5$  &$23.7$\\ 
    Efficiency curve parameters                    & $0.7$       & $0.8$      &$0.7$         &$0.8$       &$0.7$\\ 
    Efficiency curve parametrization                  & $0.7$       & $1.4$      &$1.5$         &$1.0$       &$0.9$\\ 
    $\mathcal{B}(D_{s}^{*}\to \gamma D_{s})$    & $0.8$       & $0.8$      &$0.8$        &$0.8$       &$0.8$\\ 
    $\mathcal{B}(D_{s}\to K^{+}K^{-} \pi)$                      & $3.3$       & $3.3$    & $3.3$        & $3.3$      & $3.3$\\ 
    $\mathcal{B}(D_{s}\to \kshort K)$                    & $0.8$     & $0.8$      & $0.8$        & $0.8$      & $0.8$\\ 
    $\mathcal{B}(\kshort\to \pi^{+} \pi^{-})$                        & $0.1$      & $0.1$      & $0.1$        & $0.1$      & $0.1$\\ 
    \hline 
    Total                                      &$26.3$     &$26.1$    &$25.5$       &$25.2$     &$25.4$\\
    \hline \hline
\end{tabular}
\label{tab:7}
\end{table*}

(a) The integrated luminosity is measured using Bhabha scattering events, with an uncertainty of  $1.0\%$~\cite{BESIII:lum}.

(b) The difference in the tracking efficiencies between data and MC simulation is estimated to be $1.0\%$ per track for that 
decay from IP~\cite{Ablikim:092009,Ablikim:2011kv,error_ppb2pi}. There are six charged tracks in DC-I, four tracks in DC-II 
and DC-III, and two tracks in DC-IV. The charged tracks ($\pi^{+}\pi^{-}$) of the $K_{S}^{0}$ daughters are not counted. 
The detection efficiencies of the four channels are changed by $\pm m\cdot 1.0\%$ to recalculate the Born cross section, 
where $m$ is the number of charged tracks in the channel. The larger difference 
to the nominal result, $5.7\%$, is taken as the systematic uncertainty. 

(c) The systematic uncertainty from PID is $1.0\%$ per $\pi/K$ track~\cite{Ablikim:092009,Ablikim:2011kv,error_ppb2pi}. 
This uncertainty is calculated using the same method as used for tracking efficiency, and $5.7\%$ is taken as the systematic 
uncertainty.

(d) The systematic uncertainty of the photon reconstruction efficiency is $1.0\%$ per photon~\cite{ref:gamma} based on 
studies of the processes $\chi_{c0,2}\to\pi^{0}\pi^{0},\eta\eta$. Because there are two photons in the final state, $2.0\%$ 
is assigned as the systematic uncertainty.

(e) The $\kshort$ reconstruction efficiency includes the geometric acceptance, tracking efficiency, and the candidate selection.~The difference in 
the $\kshort$ reconstruction efficiency between data and MC simulation is estimated with the control samples 
of $J/\psi \rightarrow K^{*}(892)^{\pm} K^{\mp}, K^{*}(892)^{\pm}\rightarrow \kshort \pi^{\pm}$, and 
\mbox{$J/\psi \to \phi \kshort K^{\pm}\pi^{\mp}$~\cite{ks_err}} and is determined to be 1.2$\%$.
There is one $\kshort$ candidate in DC-II and DC-III and two $\kshort$ candidates in DC-IV. The detection efficiencies of the 
channels with $\kshort$ in the final state are varied by $\pm n\cdot1.2\%$, and the resulting detection efficiencies 
are used to calculate the Born cross section, where $n$ is the number of $\kshort$ in the 
channel. The systematic uncertainty for the $\kshort$ reconstruction is estimated by the larger discrepancy between the two 
resulting Born cross sections and the nominal value.

(f) The systematic uncertainty caused by the 6C kinematic fit is estimated
by correcting the helix parameters of the simulated charged tracks~\cite{ref:kfit}. 
The result after correction is taken as the nominal one, and the difference between 
the results with and without this correction, $0.7\%$ to $0.8\%$, is taken as the systematic uncertainty.

(g) The systematic uncertainties from the $D_{s}$ and $D_{s}^{*}$ mass window requirements are due to the mass 
resolution difference between data and MC. It is estimated by correcting the mass resolution in MC to improve the 
consistency between data and MC simulation.~The correction is applied by smearing the MC distribution with a Gaussian
function, and the parameters of the Gaussian function are obtained from the control sample of 
\mbox{$\EE\to\ddstar,~D_{s}^{*\mp}\to\gamma D_s^{\mp}$}.~The mean
and standard deviation of the Gaussian function are $(0.3\pm0.6)~\mevcc$ and 
$(3.7\pm0.9)~\mevcc$ for the $\ds$, $(1.5\pm1.1)~\mevcc$ and $(0.5\pm4.2)~\mevcc$ for the $\dsstar$.~The selection 
efficiency difference with or without (used as nominal result) the smearing procedure is taken as the systematic uncertainty, 
which is $0.3\%$ and $0.1\%$ for the $\ds$ and $\dsstar$, respectively.

(h) In the nominal results, the detection efficiency and the ISR correction 
factor are obtained by assuming the Born cross section of $\EE\to\gamma X$ follows 
the line shape of the $\psi(4660)$, which is a well-established vector state in this energy region. 
Different shapes of the input cross sections result in different probabilities of generating events with the ISR photon. 
Assuming the branching fraction of $\psi(4660)\to\gamma X$ is small, the input cross section line shape is 
changed to $E_{\gamma}^{3}/s$ ($E_{\gamma}^{3}$ term comes from the radiative transition rate~\cite{ref:es} and $1/s$ is 
based on the assumption that the matrix element exhibits a similar c.m. energy dependence as the pure continuum process), 
where $E_{\gamma}$ is energy of the radiative photon and $s$ is the square of the center-of-mass energy.~There is a 
large difference between the two input line shapes, which leads to large differences in the detection efficiencies and 
the ISR correction factors.~The changes in the resultant Born cross sections are 
taken as the systematic uncertainty from the ISR correction factor~\cite{ref:weight}. 

(i) For the calculation of the Born cross section of $\EE\to\gamma X, X\to\ddstar$ with 
different $M$ under assumed $\Gamma$ and $J^{PC}$, the detection efficiency
is read from the fitted curves.~The uncertainty in the efficiency induced by the fitted parameters is 
$\Delta \epsilon_{jk} = \sqrt{(m \cdot \Delta a_{jk})^2+(\Delta b_{jk})^2+2\cdot m\cdot \mathrm{cov}(a_{jk},b_{jk})}$, where $a_{jk}$ and $b_{jk}$ are the 
parameters of the polynomial function, $\Delta a_{jk}$ and $\Delta b_{jk}$ are the 
corresponding uncertainties, and $\mathrm{cov}(a_{jk},b_{jk})$ is the
covariance between $a_{jk}$ and $b_{jk}$ given by the corresponding fit.~The detection efficiency, $\epsilon_{jk}$, is varied
by $\pm \Delta \epsilon_{jk}$, and the difference on the efficiency 
is taken as the systematic uncertainty.

(j) The systematic uncertainty from the parametrization of the efficiency 
curves is estimated by changing the fitted function from a first-order to a second-order 
polynomial function.~The average detection efficiency is  recalculated
with parameters from the alternative fitted curves.~The largest difference between the 
nominal and each fit result is taken as the systematic uncertainty. 

(k) The branching fractions of $D_{s}\to\kk\pi$, \mbox{$D_{s}\to\kshort K$, 
$D_{s}^{*}\to\gamma D_{s}$, and $\kshort\to\pp$} are taken from the PDG~\cite{PDG2018}, and 
their uncertainties are propagated to the Born cross section measurement.~The systematic uncertainties caused by the branching fractions of $D_{s}\to\kk\pi$, $D_{s}\to\kshort K$, $D_{s}^{*}\to\gamma D_{s}$,
and $\kshort\to\pp$ are $3.3\%$, $0.8\%$, $0.8\%$, and $0.1\%$, respectively.

(l)The fit-related systematic uncertainties are estimated by varying the fit models.~The largest upper limit on the number of signal events 
is taken as the upper limit of the Born cross section with the systematic uncertainty considered.~The line shapes of the background contributions 
from $\EE\to(\gamma_{\rm ISR})D_{s}^{\pm(*)}D_{s}^{*\mp}$ are taken from the MC simulations with the latest precise measurements as input. 
The uncertainties are found to be negligible. The systematic uncertainty from the fixed number of background events for the processes 
$\EE\to(\gamma_{\rm ISR})D_{s}^{(*)}D_{sJ}$ is estimated by modifying $N_{\mathrm{bkg}3}$ with $\pm 1\cdot \sigma$. ~The uncertainty arising from 
the modeling of the remaining background events is accessed by replacing the zero-order polynomial function with the line shape extracted from generic 
MC simulation. In the determination of the upper limit of the number of signal events with different $M$ under assumed width and $J^{PC}$, the mass resolutions are read from the fit curves shown in Fig.~\ref{fig:m-sigma}.~The systematic uncertainties from $m_{1}-m_{2}, \sigma_{2}/\sigma_{1}$, and $f$ are considered by varying the parameters within their uncertainties, while the systematic uncertainties from $m_{1}~{\rm and}~\sigma_{1}$, are considered by varying $\Delta m_{1}$ and $\Delta \sigma_{1}$. The largest upper limit on the number of signal events is taken, where $\Delta m_{1}$ and $\Delta \sigma_{1}$ are calculated using the same formula as used to calculate $\Delta \epsilon_{jk}$.
To take into account the systematic uncertainties in 
the upper limit calculation, the likelihood distributions are convolved 
with Gaussian functions as
\begin{linenomath*}
\begin{equation*}
    L'(x) = \int_{0}^{1}L(x; N_{\rm sig}\bar{\epsilon}/\hat{\bar{\epsilon}})\exp[-\frac{(\bar{\epsilon}-\hat{\bar{\epsilon}})^2}{2\sigma_{\rm sys.}^2}]d\bar{\epsilon},
\end{equation*}
\end{linenomath*}
where $\hat{\bar{\epsilon}}$ is the 
nominal average detection efficiency, and $\sigma_{\rm sys.}$ is the systematic uncertainty~\cite{ref:Bayesianmethod,sigma}. The upper limit of the Born cross section times the branching fraction with fixed masses and widths are listed in Table~\ref{tab:6}.~The upper limits of 
$\sigma[\EE\to\gamma X]\cdot \mathcal{B}[X\to\ddstar]$ as a function of $X$ mass
after considering all the systematic uncertainties are displayed in 
Fig.~\ref{fig:newborn}. 

\section{VIII. SUMMARY}
In summary, based on a data sample collected at $\sqrt{s}=(4681.92\pm0.08\pm0.29)~\mathrm{MeV}$ with an integrated
luminosity of $(1667.39\pm0.21\pm8.84)~\mathrm{pb}^{-1}$,  a search for $C$-even states is performed via the 
process $e^+e^- \to \gamma \ddstar$.~No statistically significant signals are observed. Upper limits on $\sigma[e^+e^- \to \gamma X]\cdot  \mathcal{B}[X \to \ddstar]$ 
\mbox{($X$= $\eta_{c2}(2D)$,~$\chi_{c1}(3P)$,~$\chi_{c2}(3P)$,~$X(4080)$,~$X(4217)$,}~and $X(4279)$) at $90\%$ C.L. are 
determined to be $10.7$ pb, $23.4$ pb, $28.3$ pb, $4.1$ pb, $13.5$ pb, and $32.4$ pb, respectively. Upper limits with other mass and width assumptions are also calculated and shown in Fig.~\ref{fig:newborn}. Furthermore, the product of the upper limit of the cross section and the branching fraction is between $4$~pb to $35$ pb.~Due to statistical limitations, the upper limit for $\psi(4660)\to\gamma\chi_{c1,2}(3P)$ is below the sensitivity of the theoretical prediction in Ref.~\cite{ref:em}. 
There is no other theoretical prediction for the cross sections of other $C$-even states.
More data samples and theoretical calculations are needed to investigate the EM transition properties of $\psi(4660)$. 

\section{ACKNOWLEDGMENTS}
The BESIII Collaboration thanks the staff of BEPCII and the IHEP computing center for their strong support. This work is supported in part by National Key R\&D Program of China under Contracts Nos. 2020YFA0406300, 2020YFA0406400; National Natural Science Foundation of China (NSFC) under Contracts Nos. 11635010, 11735014, 11835012, 11935015, 11935016, 11935018, 11961141012, 12025502, 12035009, 12035013, 12061131003, 12192260, 12192261, 12192262, 12192263, 12192264, 12192265, 12221005, 12225509, 12235017; the Chinese Academy of Sciences (CAS) Large-Scale Scientific Facility Program; the CAS Center for Excellence in Particle Physics (CCEPP); Joint Large-Scale Scientific Facility Funds of the NSFC and CAS under Contract No. U2032108; CAS Key Research Program of Frontier Sciences under Contracts Nos. QYZDJ-SSW-SLH003, QYZDJ-SSW-SLH040; 100 Talents Program of CAS; The Institute of Nuclear and Particle Physics (INPAC) and Shanghai Key Laboratory for Particle Physics and Cosmology; European Union's Horizon 2020 research and innovation programme under Marie Sklodowska-Curie grant agreement under Contract No. 894790; German Research Foundation DFG under Contracts Nos. 455635585, Collaborative Research Center CRC 1044, FOR5327, GRK 2149; Istituto Nazionale di Fisica Nucleare, Italy; Ministry of Development of Turkey under Contract No. DPT2006K-120470; National Research Foundation of Korea under Contract No. NRF-2022R1A2C1092335; National Science and Technology fund of Mongolia; National Science Research and Innovation Fund (NSRF) via the Program Management Unit for Human Resources \& Institutional Development, Research and Innovation of Thailand under Contract No. B16F640076; Polish National Science Centre under Contract No. 2019/35/O/ST2/02907; The Swedish Research Council; U. S. Department of Energy under Contract No. DE-FG02-05ER41374.

\end{document}